\documentclass{appolb}
\usepackage{graphicx}
\usepackage{subcaption}
\usepackage[normalem]{ulem}
\usepackage{comment}
\usepackage{color}

\usepackage{hyperref}

\usepackage{color}

\usepackage{xurl}

\definecolor{airforceblue}{rgb}{0.36, 0.54, 0.66}

\hypersetup{
colorlinks=true,    linkcolor={airforceblue},
filecolor={airforceblue},      
urlcolor={airforceblue},
citecolor = {airforceblue},
pdftitle={airforceblue},}

\usepackage{xspace}	

\newcommand{\AB}{\mbox{A$+$B}\xspace}
\newcommand{\pp}{\mbox{$p+p$}\xspace}
\newcommand{\dau}{\mbox{$d+$Au}\xspace}
\newcommand{\auau}{\mbox{Au$+$Au}\xspace}
\newcommand{\NN}{\mbox{N$+$N}\xspace}
\newcommand{\Lumi}{\mbox{$\mathcal{L}$}\xspace}
\newcommand{\rab}{\mbox{$R_{AB}$}\xspace}
\newcommand{\pt}{\mbox{$p_T$}\xspace}
\newcommand{\piz}{\mbox{$\pi^{0}$}\xspace}

\newcommand{\gev}{\mbox{GeV}\xspace}
\newcommand{\gevc}{\mbox{GeV/$c$}\xspace}
\newcommand{\Npart}{\mbox{$N_{\rm part}$}\xspace}
\newcommand{\Ncoll}{\mbox{$N_{\rm coll}$}\xspace}
\newcommand{\Nch}{\mbox{$N_{\rm ch}$}\xspace}
\newcommand{\dNchdeta}{\mbox{$dN_{\rm ch}/d\eta$}\xspace}
\newcommand{\Et}{\mbox{${\rm E}_T$}\xspace}

\newcommand{\Nexp}{\mbox{$N_{\rm coll}^{\rm EXP}$}\xspace}
\newcommand{\RdAuPiGl}{\mbox{$R_{d\rm Au,GL}^{\pi^0}$}\xspace}
\newcommand{\RdAuPiExp}{\mbox{$R_{d\rm Au,EXP}^{\pi^0}$}\xspace}
\newcommand{\RdAuDgGl}{\mbox{$R_{d\rm Au,GL}^{\gamma^{dir}}$}\xspace}

\newcommand{\ypidau}{\mbox{$Y_{d\rm Au}^{\pi^0}$}\xspace}
\newcommand{\ydgdau}{\mbox{$Y_{d\rm Au}^{\gamma^{dir}}$}\xspace}
\newcommand{\ypipp}{\mbox{$Y_{pp}^{\pi^0}$}\xspace}
\newcommand{\ydgpp}{\mbox{$Y_{pp}^{\gamma^{\rm dir}}$}\xspace}

\begin{document}
\title{Nuclear modification of hard scattering processes in small systems at PHENIX%
\thanks{Presented at XXIXth International Conference on Ultra-relativistic Nucleus-Nucleus Collisions}%
}
\author{Niveditha Ramasubramanian, Gabor David \\(for the PHENIX Collaboration ,\\\href{https://doi.org/10.5281/zenodo.7430208}{\tiny{\url{https://zenodo.org/record/7430208}}})}


\maketitle
\begin{abstract}
Collisions of small systems at RHIC exhibit a significant suppression of the nuclear modification factor $R_{xA}$ of jets and high momentum neutral pions in events with large event activity. This suppression is accompanied by an enhancement of $R_{xA}$ in events with low event activity. Since event activity is commonly interpreted as a measure of the centrality of the collisions, these results call into question any interpretation of the suppression in central collisions that invokes energy loss in a QGP produced small systems. In this talk, we will compare prompt photon to $\pi^0$ production measured by PHENIX in d+Au collision at $\sqrt{s}=200\,GeV$ to demonstrate that the apparent centrality dependence is not related to a nuclear modification of hard scattering processes, but likely due to deviations from the proportionality of event activity and centrality in the underlying standard Glauber model calculations. Furthermore, we will use prompt photon production in d+Au relative to p+p collisions to empirically determine the effective number of binary collisions $N_{coll}$ of a given event sample. We find that for all event selections, except for those with the highest event activity, $R_{xA}$ of $\pi^0$ is consistent with unity. By comparing p+Au and d+Au collisions, we will investigate the significance of the remaining suppression of high $p_{T}$ $\pi^0$ production in events with high event activity.

\end{abstract}

\section{Introduction}

One of the fundamental questions in relativistic heavy ion collisions (\AB) is whether particle production is the same as what would be expected from the incoherent superposition of \Lumi nucleon-nucleon (\NN) collisions from the two nuclei, where \Lumi is probabilistic, strongly correlated with the impact parameter $b$ but also depends on the initial state (IS) of the relativistic nuclei. The umbrella term ``initial state" here encompasses purely geometrical effects, like fluctuations of the shape of the nucleus and the position of nucleons therein, as well as real physics effects, like modifications of the parton distribution functions in nuclei (nPDF), gluon saturation, possible changes in \NN cross section and others. In contrast to IS, fully determined {\it before} the nuclei collide, final state (FS) effects are those processes happening during and because of the collision, and may evolve in time, like the formation of a Quark-Gluon Plasma (QGP), its transition to a hadron gas (HG), etc.  Clearly, the experimentally observed particle production is then influenced both by IS and FS.  Their joint effect on any type of particle is then characterized by the nuclear modification factor \rab defined as


\begin{equation}
\label{Eq:Rab}
R_{AB}(p_T) = \frac{(\frac{d^2N}{dp_Td\eta})_{AB}}{\langle \Lumi \rangle _{AB}* (\frac{d^2N}{dp_Td\eta})_{pp}} = \frac{Y(AB)}{\langle \Lumi \rangle_{AB} * Y(pp)}
\end{equation}

where \pt is the transverse momentum, $\eta$ is pseudorapidity, $\frac{(d^2N}{dp_Td\eta})_{AB}$ is the double-differential yield of the particle, simplified as $Y(AB)$, with similar notation for \pp.  Note that \rab doesn't differentiate between IS and FS effects.  Moreover, while $Y(AB)$ is purely an experimental observable, \Lumi is not, it is calculated using some model connecting the impact parameter to some bulk observable like charged particle multiplicity (\Nch) or transverse energy (\Et) production. Most often this mapping is based on the Glauber-model~\cite{Glauber:1970jm,Miller:2007ri}, providing the total number of nucleons interaction at least once (\Npart) and the total number of binary \NN collisions (\Ncoll).  In Eq.~\ref{Eq:Rab} the true \Lumi is replaced with the modeled \Ncoll, i.e. \rab becomes model-dependent.  The model assumes that all \NN collisions are soft (low momentum exchange).  This is true most the time, but fails in the rare cases when high \pt particles are produced at mid-rapidity.  If there are many nucleons on both sides, most \NN collisions will still be average and the Glauber-model-based mapping from event activity (\Nch) to centrality ($b$) will be correct and the calculated \Ncoll is a good approximation of the true \Lumi.  However, if on one side there are only a few nucleons, like in $p/d/^3$He+Au, and one of them participates in a hard scattering, the Glauber-mapping might become biased.  This might be a possible explanation of earlier results (like~\cite{PHENIX:2021dod}) where in $p/d/^3$He+Au at high \pt a near-uniform suppression of \piz is observed in central collisions, and, more significantly, an unexplained enhancement in peripheral collisions is seen when using \Ncoll from the Glauber-model.  Our goal is to circumvent the model, to find an experimental measure of \Lumi making \rab as defined in Eq.~\ref{Eq:Rab} a model-independent quantity, then revisit the nuclear modification in \dau collisions.

\begin{figure}[!tbp]
   \hfill
     \includegraphics[width=0.49\textwidth]{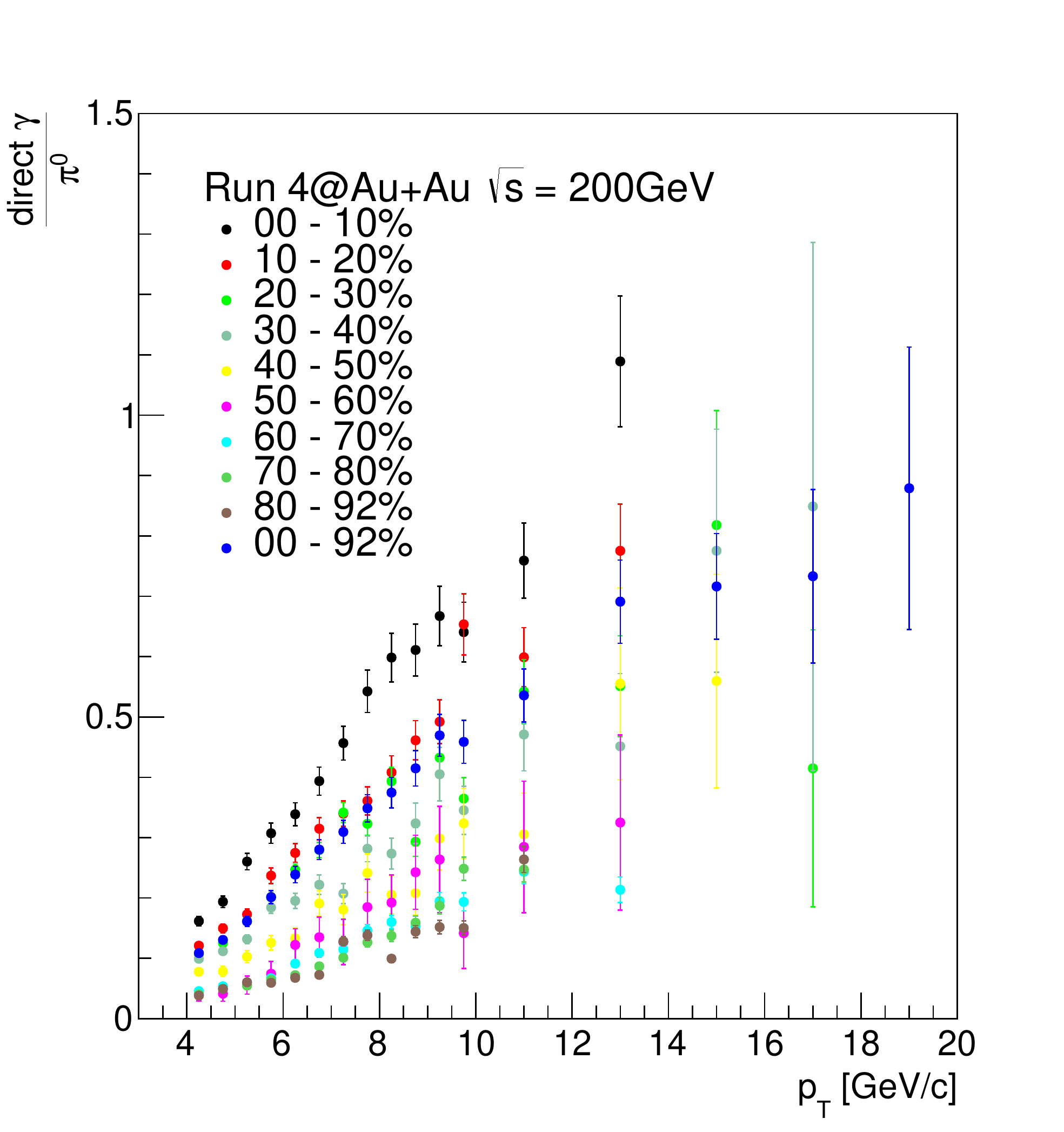}
     \includegraphics[width=0.49\textwidth]{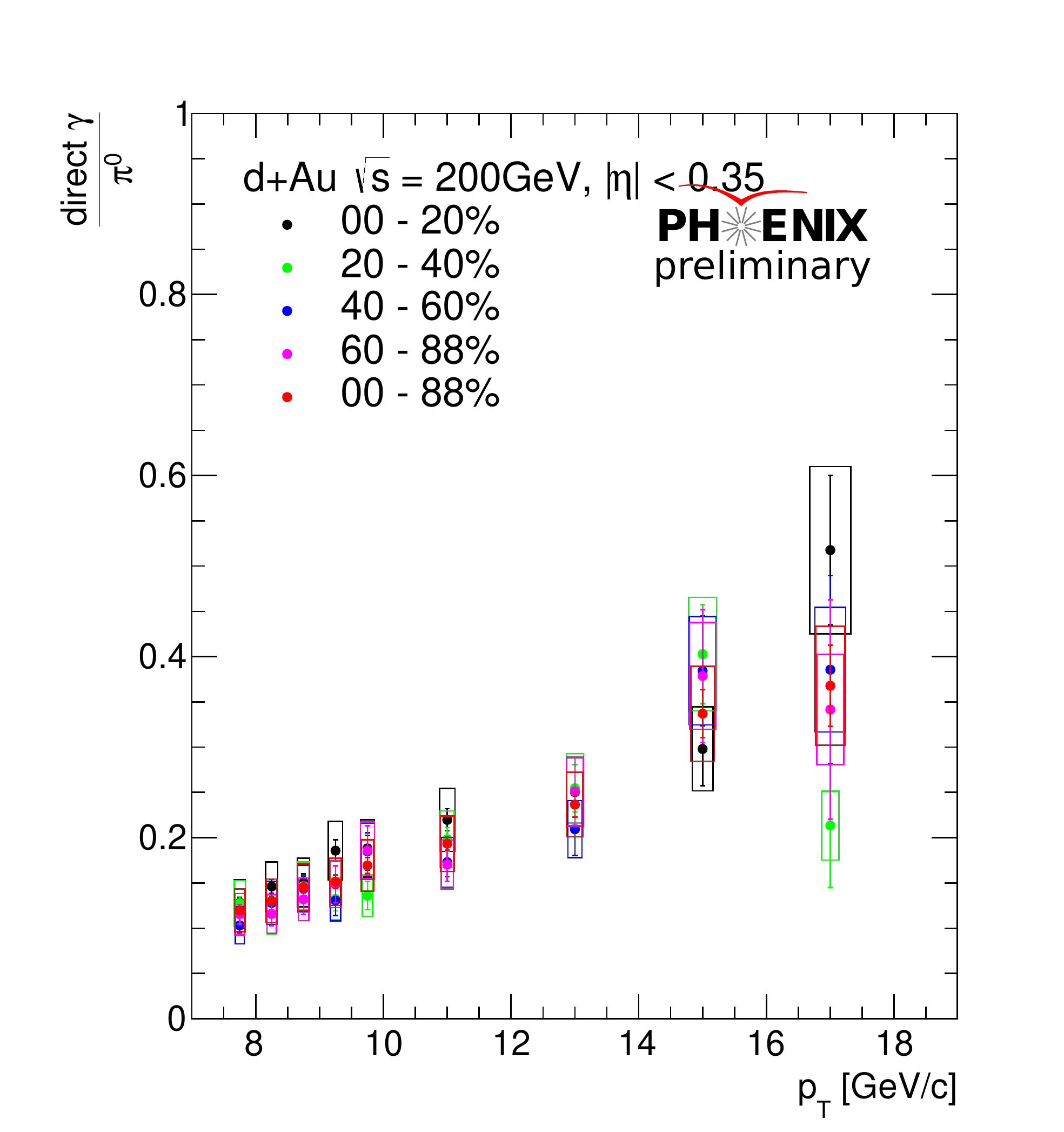}
   
  \caption{$\gamma$/\piz ratio for \auau (left panel) and \dau (right panel) collisions for all centrality selections. \auau data are from~\cite{PHENIX:2008saf,PHENIX:2012jbv}.  Note the difference in the \pt-scale.} 
  \label{fig:gammapi0}
\end{figure}

\section{Centrality dependence of the direct $\gamma$/\piz ratio at high \pt}
 Direct photons are all photons that are not coming from final state hadron decays.  At high \pt their primary source is initial hard scattering ($qg\rightarrow q\gamma$) with some small contributions from jet fragmentation~\cite{PHENIX:2012jgx}.  They are color neutral, so virtually immune to FS effects (even in the QGP their mean free path is several hundred $fm$ and increasing with $E_{\gamma}$~\cite{David:2019wpt}).  In large-on-large ion collisions their \rab calculated with Glauber-type \Ncoll is consistent with unity~\cite{PHENIX:2012jbv,CMS:2012oiv}, as expected.  On the other hand \piz in \auau collisions is proven to be increasingly suppressed with increasing event activity (centrality)~\cite{PHENIX:2001hpc,PHENIX:2008saf}.  Therefore, the direct photon over \piz ratios as a function of \pt should line up in distinct, separate bands for different centralities.  This can clearly be observed in Fig.~\ref{fig:gammapi0}, left panel.  In the right panel the same ratio is shown for \dau collisions: within statistical and systematic uncertainties the ratios overlap, they are independent of centrality.  Also, they are comparable to the ratios seen in the most peripheral \auau collisions.  Whatever physics process affects the yield of \piz at different centralities, affects the direct photons in a similar way -- contrary to \auau.

\begin{figure}[!tbp]
  \begin{subfigure}[b]{0.32\textwidth}
    \includegraphics[width=\textwidth]{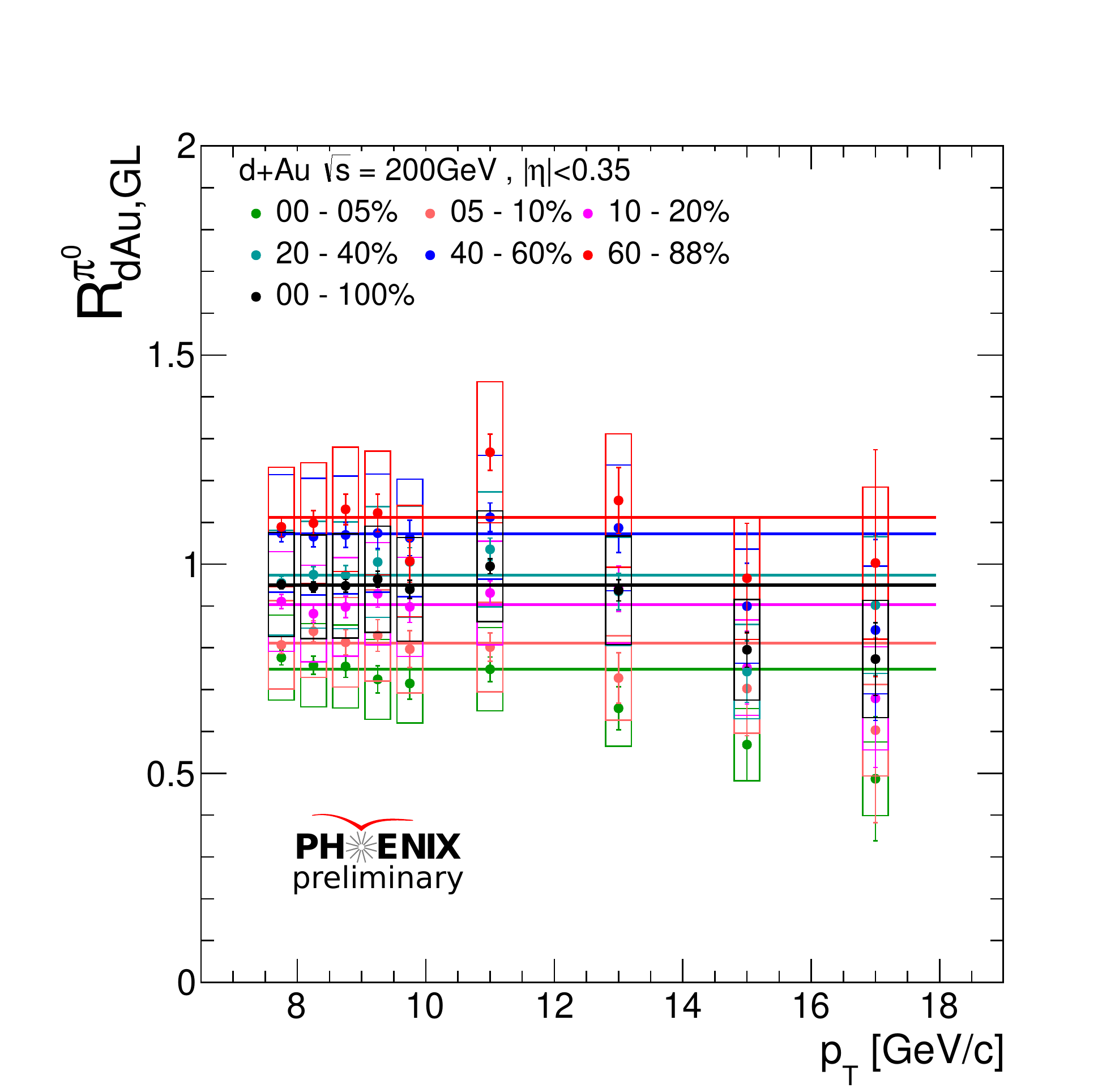}

    \label{fig:conversion_corrPi0}
  \end{subfigure}
  \hfill
  \begin{subfigure}[b]{0.32\textwidth}
    \includegraphics[width=\textwidth]{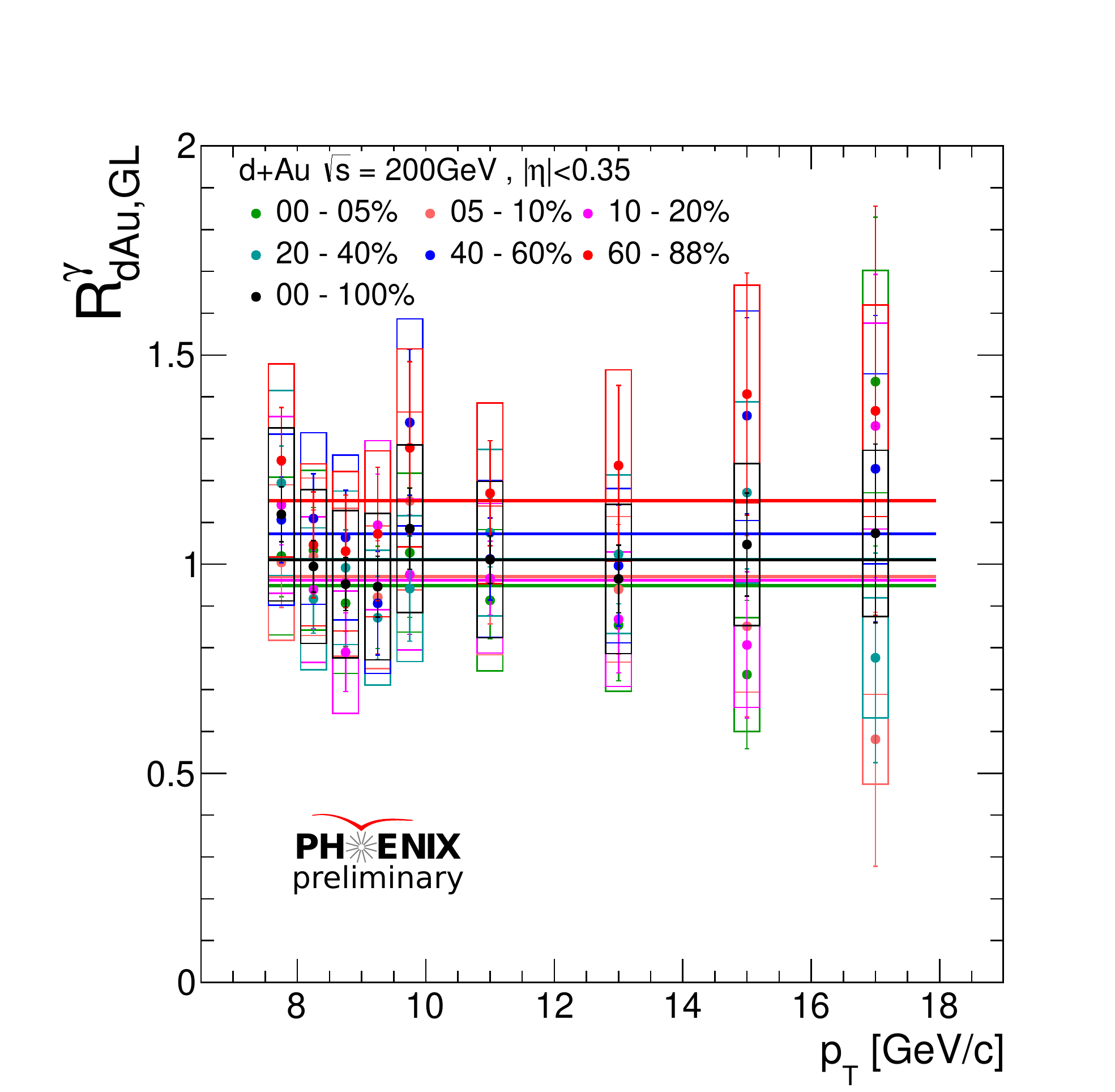}

    \label{fig:conversion_corrDP}
  \end{subfigure}
  \begin{subfigure}[b]{0.32\textwidth}
    \includegraphics[width=\textwidth]{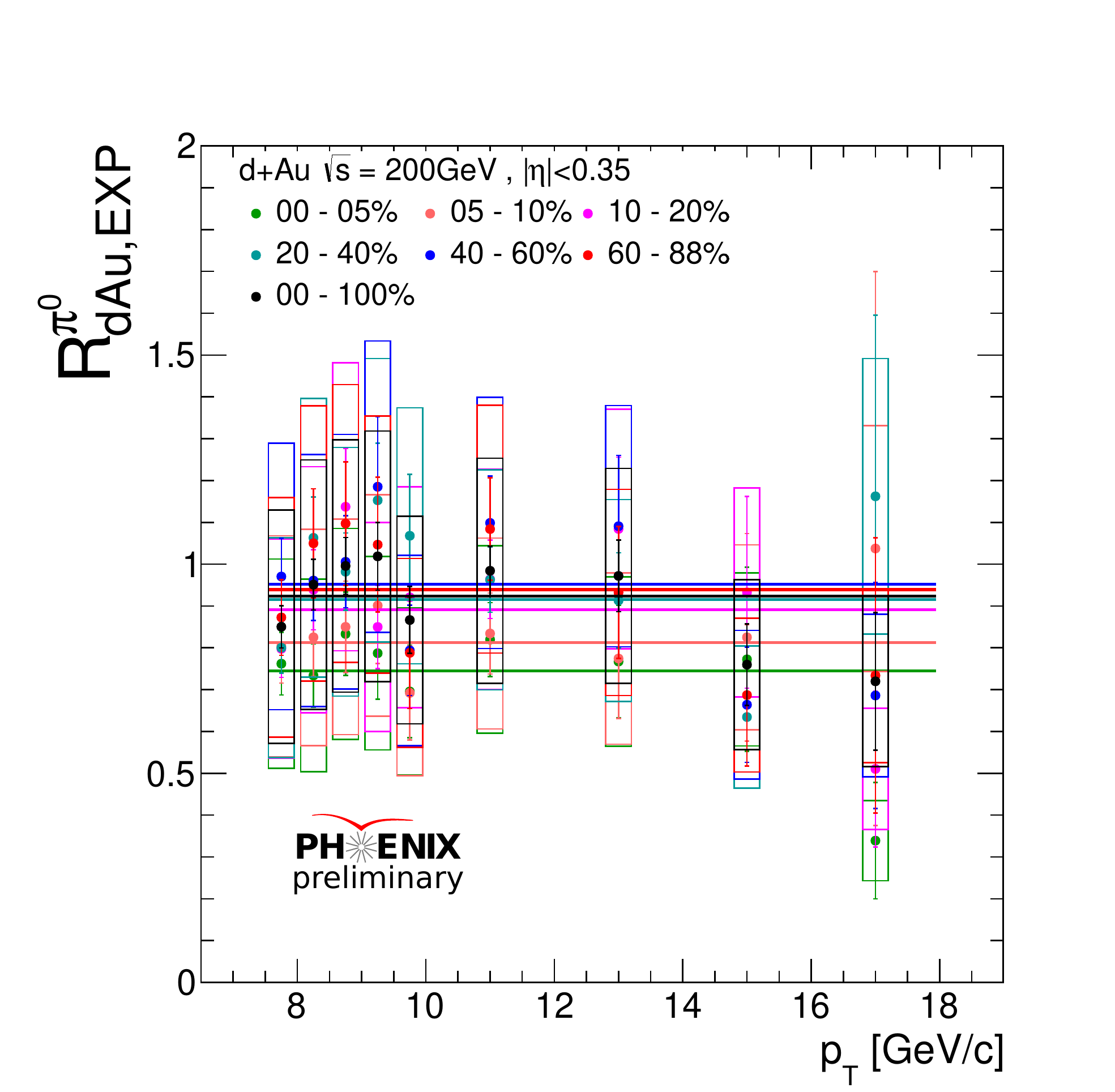}

    \label{fig:conversions_corrGP}
  \end{subfigure}
  
  \caption{$R_{dAu}$ of $\pi^0$ (left) and direct $\gamma$ (middle) obtained from $N_{coll}$ from the Glauber model and $R_{dAu}$ of $\pi^0$ (right) obtained from $N_{coll}$ derived from experiments } 
   \label{fig:rpi0}
\end{figure}

\section{Redefining the number of collisions with direct photons}

As stated above, \rab of direct photons (as in Eq.~\ref{Eq:Rab}) is unity in \AB collisions.  Turning this around one can then say that \Ncoll can be {\it defined experimentally} in the hard scattering region as the \pt-dependent ratio of the photon yields in \AB and \pp
\begin{equation}\label{eq:Ncoll_exp}
 \Nexp(\pt) = \frac{\ydgdau(\pt)}{\ydgpp(\pt)} 
\end{equation}
\noindent
and it will be a good approximation of \Lumi, devoid of biases due to FS effects.  But if this is true for large-on-large collisions, it is safe to assume that in small-on-large collisions FS effects can not be more significant.  Therefore, we propose to replace the usual Glauber \Ncoll-based  \RdAuPiGl with the \Nexp-based, purely experimental nuclear modification factor \RdAuPiExp (\pt-dependence not shown)
\begin{equation}\label{eq:double_ratio}
  \RdAuPiExp = \frac{\RdAuPiGl}{\RdAuDgGl} 
             = \frac{\ypidau/\ypipp}{\ydgdau/\ydgpp} 
             = \frac{\ypidau}{\Nexp \ \ypipp}
\end{equation}
\noindent
In Fig.~\ref{fig:rpi0} nuclear modification factors are shown using the Glauber \Ncoll (panels (a) and (b)) and with the proposed \Nexp (panel (c)), all centralities fitted with a constant.  With \Ncoll both \piz (a) and direct photons (b) show a distinct ordering from the most central (smallest \rab) to the most peripheral (largest \rab) collisions.  In panel (c) \RdAuPiExp is shown, where all centralities overlap except for the central (0-5\%, 5-10\%) bins. 

\begin{figure}
    \centering
    \includegraphics[width=0.49\textwidth]{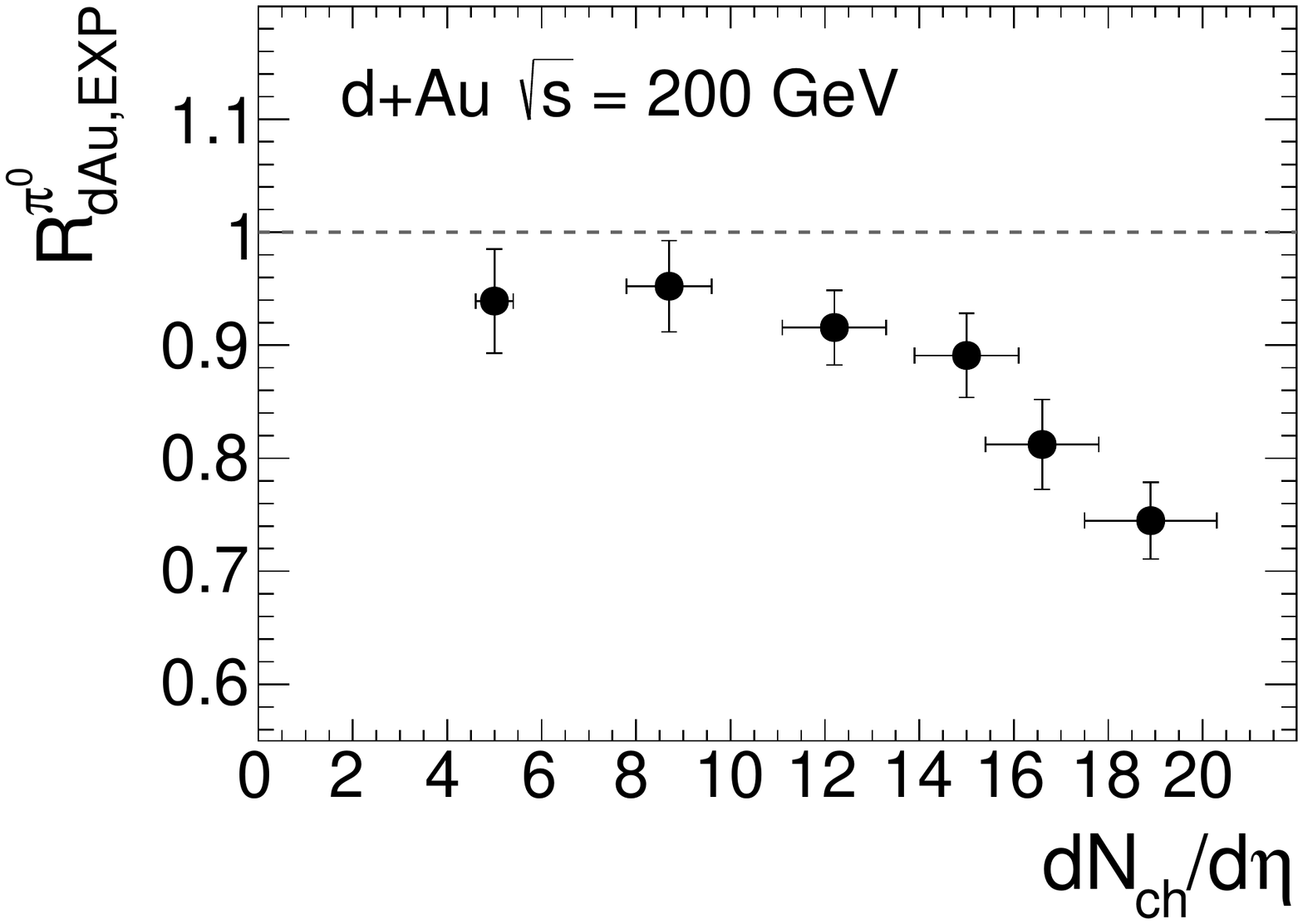}
     \caption{Average \RdAuPiExp vs event activity \dNchdeta at mid-rapidity above 8\,\gevc.  The points correspond to the 60-88\%, 40-60\%, 20-40\%, 10-20\%, 5-10\% and 0-5\% centrality bins. A global systematic error of 17.2\% is present on all the points plotted. } 
    \label{fig:rnch}
\end{figure}

The fitted values, along with their error from panel (c) in Fig.~\ref{fig:rpi0} is plotted as a function of \dNchdeta, event activity at mid-rapidity in Fig.~\ref{fig:rnch}. The points correspond to the 60-88\%, 40-60\%, 20-40\%, 10-20\%, 5-10\% and 0-5\% centrality bins.  

The systematic uncertainties are small, due to cancellations in the double ratio (Eq.~\ref{eq:double_ratio}). More importantly
there is no identified source of systematic error which has a centrality dependence, that is the systematic errors are common to all the points, therefore although the exact position of these points may be shifted up or down by 17.2\%, the relative suppression of central value with respect to the peripheral value still remains. For peripheral collisions \RdAuPiExp is consistent with unity, but in the most central ones, somewhat surprisingly, a statistically significant suppression is seen.  The reasons for that are not clear.  Based on these data alone FS effects -- parton energy loss in the small QGP droplets formed -- can not be excluded, nor can be initial state effects, or some further, second order bias in the centrality/event activity mapping.  PHENIX has collected $^3$He+Au data in 2014 and a large statistics $p$+Au dataset in 2015, which are currently analyzed for the same observables.  Seeing the evolution of the suppression with projectile size should help to clarify the situation.  Larger projectile implies larger QGP droplets, so if the suppression is due to parton energy loss, it should increase from $p$+Au to $^3$He+Au.  If  centrality bias is the reason, it should decrease with projectile size (disappearing in Au+Au) so the suppression should also decrease.

\section{Summary}
PHENIX has measured simultaneously the yield of high \pt direct photons and \piz in 200\,\gev \dau collisions to clarify the veracity of earlier claims of \piz suppression in central and enhancement in peripheral collisions.  We found that, contrary to the \auau case, in \dau the direct $\gamma$/\piz ratio is at all \pt essentially independent of the centrality selection, albeit with large systematic uncertainties.
We introduced a purely experimental measure of the number of binary collisions, \Nexp, and found that with this quantity the \piz nuclear modification factors are indeed largely independent of centrality.  However, due to the cancellation of many systematic uncertainties we could show that in the most central collisions there is still a statistically significant suppression.  The reasons of that are unclear, but ongoing analysis of the $p$+Au and $^3$He+Au will help understanding the underlying physics.


\bibliography{dauqm}   

\providecommand{\href}[2]{#2}\begingroup\raggedright\begin{thebibliography}{1}

\bibitem{Glauber:1970jm}
R.~J. Glauber and G.~Matthiae, \emph{{High-energy scattering of protons by
  nuclei}}, \href{https://doi.org/10.1016/0550-3213(70)90511-0}{\emph{Nucl.
  Phys. B} {\bfseries 21} (1970) 135--157}.

\bibitem{Miller:2007ri}
M.~L. Miller, K.~Reygers, S.~J. Sanders and P.~Steinberg, \emph{{Glauber
  modeling in high energy nuclear collisions}},
  \href{https://doi.org/10.1146/annurev.nucl.57.090506.123020}{\emph{Ann. Rev.
  Nucl. Part. Sci.} {\bfseries 57} (2007) 205--243},
  [\href{https://arxiv.org/abs/nucl-ex/0701025}{{\ttfamily nucl-ex/0701025}}].

\bibitem{PHENIX:2021dod}
{\scshape PHENIX} collaboration, U.~A. Acharya et~al., \emph{{Systematic study
  of nuclear effects in $p$ $+$Al, $p$ $+$Au, $d$ $+$Au, and $^{3}$He$+$Au
  collisions at $\sqrt{s_{_{NN}}}=200$ GeV using $\pi^0$ production}},
  \href{https://doi.org/10.1103/PhysRevC.105.064902}{\emph{Phys. Rev. C}
  {\bfseries 105} (2022) 064902},
  [\href{https://arxiv.org/abs/2111.05756}{{\ttfamily 2111.05756}}].

\bibitem{PHENIX:2008saf}
{\scshape PHENIX} collaboration, A.~Adare et~al., \emph{{Suppression pattern of
  neutral pions at high transverse momentum in Au$+$Au collisions at
  $\sqrt{s_{NN}}=$ 200 GeV and constraints on medium transport coefficients}},
  \href{https://doi.org/10.1103/PhysRevLett.101.232301}{\emph{Phys. Rev. Lett.}
  {\bfseries 101} (2008) 232301},
  [\href{https://arxiv.org/abs/0801.4020}{{\ttfamily 0801.4020}}].

\bibitem{PHENIX:2012jbv}
{\scshape PHENIX} collaboration, S.~Afanasiev et~al., \emph{{Measurement of
  Direct Photons in Au+Au Collisions at $\sqrt{s_{NN}} = 200$ GeV}},
  \href{https://doi.org/10.1103/PhysRevLett.109.152302}{\emph{Phys. Rev. Lett.}
  {\bfseries 109} (2012) 152302},
  [\href{https://arxiv.org/abs/1205.5759}{{\ttfamily 1205.5759}}].

\bibitem{PHENIX:2012jgx}
{\scshape PHENIX} collaboration, A.~Adare et~al., \emph{{Direct-Photon
  Production in $p+p$ Collisions at $\sqrt{s}=200$ GeV at Midrapidity}},
  \href{https://doi.org/10.1103/PhysRevD.86.072008}{\emph{Phys. Rev. D}
  {\bfseries 86} (2012) 072008},
  [\href{https://arxiv.org/abs/1205.5533}{{\ttfamily 1205.5533}}].

\bibitem{David:2019wpt}
G.~David, \emph{{Direct real photons in relativistic heavy ion collisions}},
  \href{https://doi.org/10.1088/1361-6633/ab6f57}{\emph{Rept. Prog. Phys.}
  {\bfseries 83} (2020) 046301},
  [\href{https://arxiv.org/abs/1907.08893}{{\ttfamily 1907.08893}}].

\bibitem{CMS:2012oiv}
{\scshape CMS} collaboration, S.~Chatrchyan et~al., \emph{{Measurement of
  isolated photon production in $pp$ and PbPb collisions at
  $\sqrt{s_{NN}}=2.76$ TeV}},
  \href{https://doi.org/10.1016/j.physletb.2012.02.077}{\emph{Phys. Lett. B}
  {\bfseries 710} (2012) 256--277},
  [\href{https://arxiv.org/abs/1201.3093}{{\ttfamily 1201.3093}}].

\bibitem{PHENIX:2001hpc}
{\scshape PHENIX} collaboration, K.~Adcox et~al., \emph{{Suppression of hadrons
  with large transverse momentum in central Au+Au collisions at $\sqrt{s_{NN}}$
  = 130-GeV}}, \href{https://doi.org/10.1103/PhysRevLett.88.022301}{\emph{Phys.
  Rev. Lett.} {\bfseries 88} (2002) 022301},
  [\href{https://arxiv.org/abs/nucl-ex/0109003}{{\ttfamily nucl-ex/0109003}}].

\end{thebibliography}\endgroup


\end{document}